# Data breaches in the catastrophe framework & beyond

*Spencer Wheatley*[1], *Annette Hofmann*[2], and *Didier Sornette*[3]

**Abstract**

Development of sustainable insurance for cyber risks, with associated benefits, inter alia requires reduction of ambiguity of the risk. Considering cyber risk, and data breaches in particular, as a man-made catastrophe clarifies the actuarial need for multiple levels of analysis – going beyond claims-driven loss statistics alone to include exposure, hazard, breach size, and so on – and necessitating specific advances in scope, quality, and standards of both data and models. The prominent human element, as well as dynamic, networked, and multi-type nature, of cyber risk makes it perhaps uniquely challenging. Complementary top-down statistical, and bottom-up analytical approaches are discussed. Focusing on data breach severity, measured in private information items ('ids') extracted, we exploit relatively mature open data for U.S. data breaches. We show that this extremely heavy-tailed risk is worsening for external attacker ('hack') events – both in frequency and severity. Writing in Q2-2018, the median predicted number of ids breached in the U.S. due to hacking, for the last 6 months of 2018, is 0.5 billion. But with a 5% chance that the figure exceeds 7 billion – doubling the historical total. 'Fortunately' the total breach in that period turned out to be near the median.

[1] Senior Scientist, Chair of Entrepreneurial Risks, ETH Zurich, Switzerland. Email: swheatley@ethz.ch

[2] Assistant Professor, School of Risk Management, Insurance and Actuarial Science, The Peter J. Tobin College of Business, St. John's University, New York, USA. Email: hofmanna@stjohns.edu

[3] Professor, Chair of Entrepreneurial Risks, ETH Zurich, Switzerland. Email: dsornette@ethz.ch

# 1. Introduction

The 3rd and 4th Industrial Revolutions have brought rapid coupling of the physical and digital (i.e., cyber) worlds, and increasing reliance on the Internet and other networked cyber-technologies. Along with great benefits, this change in *technological climate* aggravates *man-made perils* – bringing new means to commit 'old' crimes, potentially baring qualitatively new *technological risks*, and *man-made catastrophes* (e.g., see SwissRe (2018) and Chernov and Sornette (2016, 2019)). Namely, cyber technologies have ushered in rapidly evolving *cyber risk*s – e.g., taxonomized by Cebula et al. (2010), CRO Forum (2016), ENISA (2016), etc., which are placed among the most extreme in today's risk landscape (WEF, 2018) – extending crimes such as theft, fraud, extortion, business disruption, and so on to a global scope, and in some cases dramatically amplifying scale. Beyond crime, cyber-warfare also exists, with attacks on infrastructure recognized as important risks. The radical transformation in means and scale may justify treating cyber as distinct from other operational risks, or crime more generally. Proper exploitation of powerful and sometimes transformative cyber technologies justifies commensurate efforts of risk and safety analysis. However, many innovative technologies develop in an optimistic environment, leaving material risks unforeseen and perhaps externalized. It is important for 'risk thinkers' – including also academics and insurers – to insert themselves into the development, to help steer. As it stands, insurers, and society in general, struggle to grasp these somehow new cyber risks. Vendor analytics platforms on their own will not be adequate.

Here a focus is given to the insurance perspective on *data breaches.* Prominently within the class of cyber risks, *data breaches*[1] at firms/organizations can lead to theft, identity fraud, extortion, business interruption, public relations scandals, and so on. The global cost of breaches is estimated at above one hundred billion US dollars per year (Eling and Schnell, 2016b). Coverage of losses related to cyber events is typically excluded by property and liability policies. The market to cover cyber risks is growing, with insurance against data breaches forming the bulk of coverage, but coverage remains modest relative to exposure (SwissRe, 2017). As the threat worsens, development in cyber insurance can provide relief and motivate improved protection, deterrence, and mitigation through differentiated premia. Development of insurance conditional upon minimal IT security standards, adopted by a critical mass of firms, can lead to tipping points of broader improvements (Hofmann and Ramaj, 2011). As it stands, cyber insurance, risk assessment, pricing and insurance products are immature (RMS, 2017; Eling and Schnell, 2016b; Marotta et al., 2017; Romanosky et al, 2017; Shetty et al., 2018; Betterley, 2013). Early efforts to differentiate data breach risk include considering sector (Romanosky, 2016) and organization size (Wheatley et al., 2016), as well as event type.[2]

Lack of data *still* serves as a bottleneck to risk assessment and insurance pricing. And, difficulties in the insurability of cyber risks are widely acknowledged. For instance: 1) Cyber is a *quickly evolving extremely heavy-tailed* risk: reported by Maillart and Sornette (2010), and later studies (e.g., Wheatley et al. (2016), Edwards et al. (2016), Eling and Wirfs (2016)). Moreover, as illustrated by Eling and Schnell (2018), cyber insurance portfolios may lead insurers into a *non-diversification trap*: where the tails are sufficiently *heavy and dependent* that attempted diversification is ineffective or even counterproductive (e.g., Bouchaud et al., 1998; Ibragimov et al. 2007; 2009). 2) Lack of information and standards cause general risk ambiguity as well as asymmetric information and adverse selection problems in underwriting. 3) The important and challenging-to-model role of human behavior on both the attacker and defender/victim sides.[3] And, 4) risk inter-dependencies mean that the security level of individual firms affects others (Kumar et al., 2007), leading to so-called Interdependent Security (IDS) game-theoretic issues (Kunreuther and Heal, 2003).

---

[1] Of basic personal information 'ids'—such as name, social security number, address, etc., as well as accounts, transactions, and privileged communications. Article 4(12) of the GDPR defines a personal data breach as "a breach of security leading to the accidental or unlawful destruction, loss, alteration, unauthorized disclosure of, or access, personal data transmitted, stored or otherwise processed."

[2] E.g., software versus network based events (Boehme and Schwartz (2010), Oeguet et al. (2011), Mukhopadhyay et al, 2013); and data loss or unauthorized modification, data breach, and financial fraud (Cebula et al. (2010), Biener et al. (2015), Eling and Schnell (2016)).

Many have mentioned that cyber risks are like *nat-cats* (natural catastrophes – such as floods, earthquakes, or huge fires), at least in terms of scale. More precisely, they belong in the class of *man-made cats*. E.g., in CRO Forum (2018), cyber risks are posed as the technological equivalent of extreme weather. SCOR (2017) and Stalder (2017) mentioned the potential to learn from *nat-cat* modelling when analyzing cyber and other man-made catastrophe risks. Note that each *cat* type has a different set of phenomena that need to be understood in detail. Cyber is of a somewhat novel nature.

Here we formally analyze cyber risks, and data breaches in particular, as *cats*, according to the framework typically applied to *nat-cats* in an insurance context (Grossi and Kunreuther, 2005). Indeed a single level of analysis, such as financial loss, is limited. For instance, as breach mitigation practices improve, the financial loss of a given breach will be reduced. However, breaches may remain large and frequent, and generally undergo distributional changes that will be difficult to identify on the basis of loss statistics alone. As a consequence, technical measurements determining exposure (size, type, and system structure), and vulnerability ('people, processes, and technology') should be used in addition to claim history. A cat analysis synthesizes the essential elements, helping clarify the current state of data and modelling, as well as how to improve. A broader discussion of top-down statistical risk assessment with bottom-up safety analysis is offered to help guide future developments in modelling and insurance risk assessment. Drawing from experience in the nuclear safety sector, inter alia, we emphasize the importance of learning from *near-miss* events, which tend to be ignored by cost-level analysis alone.

Addressing the dynamic aspect of cyber risk, comprehensive analyses indicate that the risks are worsening: Europol (2016) provided that "[t]he volume, scope and material cost of cybercrime all remain on an upward trend and have reached very high levels". Moreover, global ransomware losses were estimated to exceed USD 5 billion in 2017, up 15 fold from 2015. Next, RSA (2018) stated that, in 2017, account takeover attacks tripled, resulting in over $5 billion in associated losses. And, Proofpoint (2019) found that, relative to the previous fiscal quarter, email-based corporate credential phishing attacks quadrupled and web-based social engineering attacks tripled. Considering data breaches, for U.S. firms with severity measured in private information items ('ids') extracted, Wheatley et al. (2016) found that data breaches have become heavier tailed, whereas Edwards et al. (2016) argues that they have not. Here this problem – focused on the 'damage' level of the cat framework – is revisited with updated data. We distinguish external attacker ('hack') events as a worsening risk class, offering a step towards overall characterization of data breach risk.

## 2. Cyber risk data breaches in the cat framework

We frame data breaches as a *cat*, which helps clarify the current state of cyber risk data and modelling – largely in terms of a statistical approach – as well as how to improve. This synthesis is done at high level, where more exhaustive references are in reviews by Marotta (2017) and Eling and Schnell (2016b). In principal, an understanding of all of the following levels is necessary to characterize the risk. As we focus on data breaches, we exclude more severe broader consequences of cyber catastrophes, such as crisis and disruption due to a cyber-attack on critical infrastructure or other large accumulation scenarios foreseen to cost in excess of 1 percent of the GDP of affected regions (Lloyds, 2015; SCOR, 2017). Fortunately, such a cyber-catastrophe has not yet occurred, and the frequency of such events is deeply uncertain.

[3] See Gordon and Sohail (2003), Baer and Parkinson (2007), Opadhyay and Rao (2009), Shackelford (2012), Mukhopadhyay et al. (2013), Eling and Wirfs (2015) as well as Eling et al. (2016a, 2016b).

## 2.1 Cat modules

**2.1.1 Exposure:** *The quantity, type, structure, and value of sensitive information at risk at a given firm.* Different risk-classes of information types exist. Further, this mass of data needs to be mapped onto the IT systems that host them, to specify component and system level exposure potentials. Insurers can solicit this information from clients, building databases over time. Academic studies suffer from a lack of such specific exposure data. Wheatley et al. (2016) inferred that extreme breach risk scales with the size of the largest sub-unit in an organization. However, more specific and direct analysis remains desirable.

More digital data is available worldwide, with more users connected each day, and large databases (often web-based cloud services) concentrate pools of information. 50% of the current global population are already Internet users. Basic personal information includes identity (e.g., name, address, date of birth, etc.), account information (e.g., web services, email, financial), and medical secrets (even genetic information). Broader information includes personal (e.g., political view) and transactional secrets. To give an impression, each minute there are: 3.8 Million Google queries, 188 million emails exchanged, 4.5 million Youtube videos viewed, and hundreds of thousands of images shared[4]. Microsoft[5] estimates that data volumes online will be 50 times greater in 2020 than in 2016. Cisco estimated that 95% of data center traffic will occur on the cloud by 2021[6]. A firm should maintain sharp awareness of the data that it processes and stores, as well as its structure, and people/processes interacting with it.

**2.1.2 Hazard ('threat'):** *Intensity of diverse events that may lead to data breach, including intentional and accidental events, as well as technical and/or human causes.* As both intentional attacks and unintentional disclosures are considered, the hazard spans both security and safety/reliability. Attackers – so called *threat actors* – can be internal, external, and state actors cannot be excluded. Diverse means/modes of attack exist ('vectors') such as the Internet, private networks, social engineering attacks, theft of hardware, and other strategies. Further, there have been many large breaches due to unintentional disclosures – proving the important role of basic human error in current systems. Characterizing hazard consists of identifying the distinct (evolving) types, their frequency, and perhaps a strength/intensity measure for a given hazard type which may occur on different scales.

Cyber-criminals are highly motivated by lucrative profits of e.g. financial theft, identity fraud, and blackmail/extortion, and a relatively low risk of getting caught. Further, open software packages lower the level of entry into such activities. Wheatley et al. (2016) showed that the attack hazard is higher for larger organizations. There are many measures of the hazard/threat intensity – e.g., Proofpoint (2019) find that 83% of surveyed respondent firms experienced a phishing attack in 2018, a 76% increase over the previous year. Although all attempted attacks are difficult to identify, such statistics along with qualitative information about the rapidly evolving techniques used by cyber criminals is essential to assess threat level, and effectiveness of protection measures. Ultimately, combat, control, and deterrence of the hazard is a state function, with inter-state cooperation necessary to effectively address this global phenomena (e.g., https://cyberstability.org/). However, the state, insurers, and firms can monitor emerging risks, e.g., where new techniques are first being tested on smaller victims. This information can be gleaned from police and insurance reports, surveillance intelligence, or through other event reporting mechanisms. In principle, such information would enable dynamic risk management – spanning assessment, protection, and effect mitigation. It is understood that some states and insurers are pursuing such directions.

[4] https://www.weforum.org/agenda/2019/03/what-happens-in-an-internet-minute-in-2019/
[5] https://www.microsoft.com/security/blog/2016/01/27/the-emerging-era-of-cyber-defense-and-cybercrime/
[6] https://www.zdnet.com/article/cloud-computing-will-virtually-replace-traditional-data-centers-within-three-years/

**2.1.3 Vulnerability:** *Susceptibility to breach, measuring the damaging effect of hazard events on exposures.* In cat modelling, *vulnerability curves* map hazard intensity to degree of damage. For each combination of exposure and hazard type and intensity, the probability of breach, and distribution of damage in event of breach, are the key objects. It remains to be seen how much variability can be explained through exposure and hazard factors. Ranging from common passwords, to shared systems (including data pools), software and vendors, both exposures and vulnerabilities of firms are interdependent – potentially with global scope. Insurance consequences of this are well known: a reduced risk pooling, suboptimal economic outcomes, and a potential for relief from reinsurance market development. The pitfall of missing common modes of failure should be avoided[7].

Relative to attackers, many victims are complacent[8] and unsophisticated in IT security. In addition to the trend for growing amounts of data being loaded onto networks, and concentration of data pools on 'the cloud', use of mobile devices and 'smart/IoT' networked appliances lead to mushrooming of the *attack surface.* E.g., Gartner (2018) "[s]ays 8.4 Billion Connected "Things" Will Be in Use in 2017, Up 31 Percent From 2016". This surface is very porous as, in addition to widespread unsafe IT practices of end-users, the software business model allows for the sale of buggy software (Chernov and Sornette, 2016), which the inertial legal and insurance apparatus struggles to keep up with. It is sobering to see the volume of vulnerable devices connected to the Internet (e.g., with open ports), identified by shodan.io.

Management of vulnerability is the realm of *protection*, where firms/organizations themselves, along with help from technical support organizations and regulators, fight an ongoing battle against cyber threats. For instance, the U.S. government facilitates a vulnerability database (https://nvd.nist.gov), and developed the Common Vulnerability Scoring System to measure the severity of vulnerabilities, based on qualitative-technical characteristics. The EU Cybersecurity Agency certifies products and services along with general data protection standards ('GDPR' – see EDBP (2019)), and standards for maturity of IT security exist (e.g., ISO/IEC 27001). As for *exposure*, insurers can play a constructive role here through *rigorous* IT security audits, with credit reflected in premia. Worth noting is that 'hackathons' and continued self-attack as *stress test* are used, and will likely become indispensable for firms, insurers, and regulators – going beyond superficial 'IT hygiene' tests. In comparison with nat-cats, where one largely relies on models and simulation to evaluate response of assets to major natural perils, cyber-attack scenarios can be played out to a large extent, as a test/audit, without being prohibitively costly or destructive. Considering scenarios is clearly a good practice to assess and manage vulnerability. For instance, Lloyd's (2017) considered a hack of a cloud service provider (which they appropriately called the "cyber equivalent of a hurricane"), as well as exploitation of a mass software vulnerability. However, Kessler (2018) found that only 25% of surveyed Swiss firms have undertaken "modelling of potential cyber loss scenarios".

**2.1.4 Damage:** *The amount and type of data breached within a breach event.* This is the measure of the realized data breach, determined as the stochastic response of a given vulnerable exposure to a hazardous event. Each event is measured in terms of breached identity items (*ids*), and distinguished by id type to allow differentiation of damage. For data breaches, it is natural to consider *complete breach events* – neither disaggregates, such as damage incurred by single affected individuals, nor aggregates, e.g., over time. Event definition is clear when the breach is confined to a single organization, but is complicated by interdependencies: E.g., a breach due to a common vendor can be treated as a single event, or a set of dependent events at affected clients, with the latter being a richer description. Another complication are secondary effects, which can serve as a multiplier: i.e., damage due to the exploitation of information combined across multiple breaches. E.g., Troy Hunt posted a collection of 773 million ids collected across multiple breaches[9]. Aside from such difficulties, this measure of damage is objectively defined, typically reported, and independent of the (also evolving) factors that map the breach size to cost.

---

[7] Such as the concentration of semiconductor manufacturing in a region of high flood risk in Thailand, e.g., see Romero, J. "The Lessons of Thailand's Flood", IEEE Spectrum, 1 Nov 2012. Also: in November 2017, it was confirmed that millions of Intel chips have a major vulnerability, potentially allowing arbitrary remote code execution and privileged information access. [https://www.us-cert.gov/ncas/current-activity/2017/11/21/Intel-Firmware-Vulnerability].

[8] Based on 75% of worldwide browser traffic on the Internet, Maillart et al. (2011) document a massive "law of procrastination" in the form of a general class of power law behavior of the waiting times to update outdated software. See also Saichev and Sornette (2010).

Following the EU Cybersecurity Act[10], a processor of personal data must report any data breaches within 72 hours if they have an adverse effect on user privacy, with sanctions enforced ca. 2018. All 50 U.S. states have implemented mandatory data breach reporting. Using these data, Maillart and Sornette (2010) quantified the extremely heavy tailed nature of data breach sizes, shown to be increasingly heavy-tailed by Wheatley et al. (2016), but also with a highly consequential and significant upper truncation. This is natural, given that the largest possible breach is determined by the largest existing exposure. It was further shown that *extreme* breaches had an increasing relationship with victim organization size, which has been supported by other studies. However, large firms also suffered small breaches – showing the importance of adequately capturing breach size variability for a given firm size, and not reducing to mean-value / deterministic vulnerability curves.

**2.1.5 Loss:** *The relevant measure of negative consequences (e.g., economic or full social cost), or liabilities in a contract.* We consider (combined first and third party) losses consistently with the literature, acknowledging that a full social cost is difficult to estimate. Relying on historical insurance claims is questionable due to inconsistent and typically incomplete cover. The key object here is a stochastic mapping from damage to loss, conditional upon other factors describing the victim and breach event. E.g., if the breach occurred over a longer period, the attacker would have had more time to exploit the stolen information. Further, agile and effective breach response measures mitigate losses (Ponemon, 2018). Dependence of cost on country, sector, event type, and so on have been studied (Ponemon, 2017, 2018; Eling and Wirfs, 2015; Romanosky, 2016).

Based on comprehensive interviews with firms, IBM/ Ponemon publish annual reports on the cost of data breaches – with a current global average of $148 per id (2018) for breaches of *moderate size* (up to 100'000 ids). Emphasis is often put on such average costs, despite heavy-tailed distributions, where, as noted by NetDiligence (2014), single point values are misleading. Using Ponemon's (not public) data, again for *moderate breach sizes*, Jacobs (2014) posed a sublinear regression, where cost scales with breach size to the power of 0.76, with an observed range in cost of about a *factor of five* for any given breach size. A sublinear relationship is natural, as the large breaches are more visible, the exploitation of larger breaches takes a long time and is hence easier to mitigate. As positive example – in the face of a worsening hazard – Eling & Wirfs (2019) have inferred that resultant costs are not worsening, presumably due to better mitigation. Romanosky (2016) studied cost-size relationships using the (not open) Advisen data, including *larger breaches,* quantifying a sublinear cost-size relationship with power 0.3. With 600 breaches with mean and max sizes of 5 Mil., and 572 Mil. ids, and mean and max costs of $5 Mil. and $572 Mil, this yields a mean of about $1 per id for breaches for *large breaches*. Ponemon (2018) considered large breaches – surveying 11 firms experiencing breaches between 1 and 50 million ids to disclose their loss – also finding cost per id decreasing with size, down to about $7 per id at the 50 million id size. In approximate agreement with the

[9] "Collection #1". https://www.troyhunt.com/the-773-million-record-collection-1-data-reach/
[10] See https://www.enisa.europa.eu/news/enisa-news/european-commission-proposal-on-a-regulation-on-the-future-of-enisa.

above, it is interesting to note that diverse media reports[11] for *extreme breaches* (in excess of 50 Mil. items) place cost in the ranges between $1 and $5 per id. This is far from a proper characterization of variability, but provides a heuristic where say about 80 percent of events should have per capita costs within that range. Detailed analysis of the development of breach costs, especially large ones, as a function of breach response/mitigation characteristics will be highly attractive. The full cost of breaches, including externalities / social costs, should also be carefully examined – noting the GDPR regulation allows for fines of up to 4% of annual revenue.

## 2.2 Current breach data overview & total cost

According to SwissRe (2018), in 2017 total economic losses due to natural and man-made catastrophes[12] were $337 billion, with the man-made contribution forming the minority. The total socio-economic cost of a (*rare)* major nuclear accident is also estimated to be on this order (e.g., see Sornette et al. (2019)). Now, we review the frequency and severity of extreme and moderate data breach. To estimate absolute costs, it is useful to extrapolate cost information from a subset onto the broader sample of breach sizes – as done by Edwards et al. (2016), Romanosky (2016), and reviewed by Eling & Schnell (2016b).

To briefly review major data sources: the widely used PRC database (https://privacyrights.org/data-breaches) collects breach events in the USA, and contains 8'754 events with breach size (# ids) totaling 11.6 bil. items, counting until 1-May-2019. Risk Based Security maintains a proprietary global database, that we have not been given access to. Publicized summary statistics include: 39'917 events up to 1-May-2019, totaling 27 bil. ids. Further, 10% of that total occurred in Q1 2019 across 2'504 breaches! The Risk Based Security total of near 3-fold that of the PRC, if attributed to global scope, is approximately consistent with the ratio of US GDP to global GDP. It is also clearly apparent that the Q1 2019 data from PRC is highly incomplete, due to delay in reporting and inclusion there. The proprietary database of Advisen claims to have in excess of 75'000 events, including both size and cost information. According to Eling and Wirfs (2019), the SAS OpRisk database contains 1'579 breach related loss events in excess of $100'000, with a mean and median losses of $43 mil. and $1.5 mil. However, breach size information was not considered.

Such databases are incomplete, especially for smaller events. E.g., the European Data Protection Board got 64'684 GDPR applications related to data breaches in its first 9 months (EDBP, 2019). Incomplete and varying reporting behavior is a standard problem that can be addressed by representative surveys, allowing for population level inference: Ponemon's (2018) survey estimated that 28% of firms suffered a data breach of *moderate size* (less than 100'000 ids) in 2018. KMPG (2016) found that 60% of surveyed small UK businesses had experienced a breach at some time, with 63% of those having suffered a breach within the previous year – indicating a higher breach rate of 38% of firms in 2015. Esentire (2019) found 44% of the 600 surveyed firms had at some point suffered a significant breach *due to a vendor.* To identify the order of magnitude of cost of moderate breaches, we apply the 0.28 probability of breach per firm per year to all US firms with more than 100 employees (about 109'000 of a total of 6 mil., according to 2015 US Census). Setting the average moderate breach size to 10'000 at a cost of $150 per item, gives $42 bil. annually for moderate breaches in the US. Further study of the annual breach probability, differentiated by size, sector, and so on, is highly desirable. Assuming that the data in the PRC database is more complete for *large breaches* (size above 100'000 ids), taking all such *data breaches* at US firms and, applying $5 per id gives $58 bil. for the full history, 85% of which was accumulated in the past 5 years – averaging to $10 bil. per year for the US.

[11] https://www.csoonline.com/article/2130877/the-biggest-data-breaches-of-the-21st-century.html
[12] Among other conditions, with event loss in excess of $20-50 mil., depending on sector.

Together, the moderate and large breach costs for the US plausibly indicate a global cost on the order of $100 bil., approximately in agreement with the estimates reviewed by Eling & Schnell (2016b).

## 3. Cyber data breaches: risk assessment & safety/security analysis

We now address two different objectives and activities, both of interest to the insurer, which we label as: 1) risk assessment: striving for best estimate of the absolute risk of *data breach* at a given firm on the basis of statistics, and 2) safety/security analysis: balancing *overall* safety/security with profit in design and operation.

Putting cat-modelling aside, general cyber risk, and data breach as a relatively simple subset of it, are *man-made*, and taking place in socio-technical systems with important cyber elements. It is worth reflecting on analysis of safety (relating to unintentional harm) and security (relating to intentional harm) in such systems: One can draw superficial comparison of data breach to industrial systems that store, process, and potentially breach hazardous materials. But, the networked cyber element is another defining feature found in ICT systems. Further, the risk of cyber systems are likely more driven by security than safety concerns. Safety/security analysis of 'cyber' is challenging inter alia due to speed of evolution (of risk and system), interdependency, difficulty in assessing human and software reliability, and pervasive and diverse use – making control and high standards difficult. In such cases the *resilience framework*[13] (Kovalenko and Sornette, 2016; Trump et al., 2018) is attractive – here emphasizing detection, response, and recovery to disruptions, as opposed to singularly focusing on hardening protection – e.g., because effectiveness of resilience does not necessarily depend upon the specific disruption being foreseen. It is hence promising for insurers to encourage and credit *resilience*, where Ponemon (2018) has quantified effects of data breach response and recovery characteristics. In the cat framework, *resilience* would feature in both vulnerability mappings onto damage, but also the relationship between financial consequences and breach size.

Safety/security analysis remains important, for which frameworks and tools exist to identify potentials for harm and their causalities, with cyber-specific methods called *threat modelling* (Shevchenko et al., 2018) – which are complementary (with none sufficient on their own). For risk assessment, to enrich and go beyond pure statistical observation, as well as to inform safe design and operation, a *socio-technical system model* is needed. For data breaches specifically, this requires modelling 1) the *focal system*: data storage and processing elements, flows, and boundaries, 2) its *control system*, including sensors, actuators, and control processes, and 3) corresponding *human and organizational structure and processes*. Common exposure and vulnerability (data, hardware, and software) across firms need to be mapped, or imposed at an aggregate level. Safety/security must be considered under normal operation, threat, accident conditions, and post-accident response. For example, the framework of Leveson (2011) can help rapidly provide a comprehensive 'systems view', and identify high level severe vulnerabilities and causalities of unsafe states. The cyber-specific STRIDE framework goes in this direction where, in sophisticated cyber systems, the number of identified threats becomes very large.

In terms of statistical risk estimation, such high level system models should already provide important technical and organizational factors to differentiate risk – at exposure and vulnerability levels in particular. Broad and well-structured operating experience, appropriately pooled, may allow for more granular statistical risk quantification. Additionally, subjecting a system (or its model) to test scenarios, with varying degrees of surprise and invasiveness ranging from survey to unannounced test attack, can serve to further differentiate system-level risk.

[13] There is no commonly agreed definition of resilience: In general, it is the ability of the system to sustain or restore basic functionality following a risk source or event (even unknown) – SRA (2015). A concrete definition of resilience for critical infrastructure has been given by Kröger (2019).

Regarding safety/security, optimization and refinement require more detailed modelling, and various analyses are aided by quantification. Traditional methods, based on decomposition and causality, like *logic (event and fault) trees, human reliability analysis,* and *correlated failure* methods have limits in capturing dynamic behavior and can become prohibitively laborious for full detailed analysis of sophisticated cyber systems, but nevertheless aid identification of safety issues and inform operation at appropriate levels. In particular, such *bottom-up* classical approaches should not be expected to fully capture/quantify absolute risk. It is profitable to consider lessons from the mature and pioneering use of such approaches in nuclear safety, and from the challenges experienced in quantification of absolute risk, combined with limited experience of 'breach' accidents (Sornette et al. 2018). Top-down risk estimation and bottom-up safety analyses are complementary and, in critical instances, should be done in the greatest possible detail.

For instance, as illustrated in the 'STAMP' model of Leveson (2011), technical and regulatory designs will co-evolve with operating experience, with incident and accident reporting as an important component. As such, the information stream should be made as rich as possible. Further, quantification in cyber systems relies on statistics derived from operating experience, which also provides further information about the peril (e.g., new hazards, vulnerabilities, etc.). Data breaches are a member of the class of operational risks, for which the benefit of sharing experience of relatively rare severe operational losses has been recognized and led to anonymized sharing of events in aggregate databases (e.g., SAS OpRisk, and ORX). However, the special nature of cyber may justify a separate treatment and structure, e.g., following the cat-modelling discussion. Existing operational risk databases are based on voluntary reporting, anonymity, and sector-specific focuses. E.g., the ORX database reports only $4.2 bil. in global operational loss of $4.2 bil. between 2012 and 2017, whereas Eling and Wirfs (2019) find a total of nearly $70 bil. in losses related to data breach in the OpRisk database. Neither claims to be complete nor representative – making inference to broad populations difficult.

It is informative to consider the reporting of operational events and experience within the nuclear sector (e.g., see Sornette et al. (2018)). Plant operators voluntarily contribute detailed information in the confidential World Association of Nuclear Operators (WANO) database, which is understood to be useful for component reliability estimation. The International Atomic Energy Association (IAEA) introduced the International Nuclear Event Scale (INES) as a relatively simple 8 point scale of escalating severity for incidents and accidents, reported globally. Operators are obligated to report incidents and accidents with INES score, however the IAEA has cited a lost learning opportunity due to absence of global sharing of 'low level event' incidents, and near-misses / *accident precursors.* In particular, a 0-1 raw outcome measure is reductive where an *incident* (e.g., a latent vulnerability, or a failed significant attack) could have come arbitrarily close to having been an *accident (e.g., a breach). Precursor analysis* (Kunreuther, et al, 2004; Ayoub, et al., 2019) quantifies the probability of a near-accident having been an accident, using probabilistic tree-based methods in nuclear. As an example, the Fukushima Daini nuclear power station was hit by the 2011 Tōhoku seaquake and tsunami, but happened to retain an external power line, and hence did not fall into complete blackout, unlike the less fortunate Daichi site. Was Daini 'better', or 'lucky'?

Similar ideas could prove useful for cyber incidents, immediately showing that the OpRisk approach of imposing a minimum financial loss for reported events can exclude interesting events (precursors). In fact, there are many reported 'data breach' examples where, e.g., a large amount of data was left exposed but by chance was not exploited (see PRC database), whereas almost-successful attacks seem less frequently reported. Assessment of significance of cyber precursors can be done on the basis of probability of damage in similar circumstances, or estimated using probabilistic tree-based methods, as in nuclear – even if only on a highly uncertain (order of magnitude) basis. The value of precursor analysis for insurers, operation, and regulation is clear. It is hence promising that the CRO Forum (2016) questioned how near-miss thresholds should be defined, and some GDRP guidelines mention reporting of near-misses. Simple and clear guidelines and examples will be necessary for broad and consistent reporting practice. In 2017, there was an OECD workshop on organizing cyber incident reporting and sharing[14]. To briefly address this problem in view of the above: 1) Data is needed at all levels of the (nat-)cat framework. 2) Regulatory/legal compulsion will be required for broad and consistent reporting. 3) Something analogous to the nuclear INES – being simple, rapid, approximate, and omitting confidential details – could strive for complete application and would support risk analysis, benefitting from academic involvement. 4) (Re-)insurers can pool claims experience, allowing for more sophisticated analysis of events raising claims. 5) Detailed and potentially more sensitive exchange could take place under stronger confidentiality, and perhaps away from insurers and regulators, for instance on the model of the World Association of Nuclear Operators (WANO). This could support bottom-up safety objectives.

# 4. Data breach damage in the US

We now consider the cat framework *damage* level alone for US events. We do not normalize for exposure (as done in Wheatley et al. (2016)), vulnerability level or hazard level. We use the PRC data[15] merged and harmonized with events from the former Open Security Foundation Dataloss Data Base, resulting in a set of 1’713 breach events occurring to organizations – including businesses, government, educational institutions, and healthcare institutions – with event severity in excess of 10,000 ids. These events correspond to breaches occurring between 01/2005 and 09/2017, and were collected in Q2-2018. To our knowledge, this is the most comprehensive open scientific dataset for breaches, covering social security number, credit/debit card number, email/password/user name, protected health information, driver’s license, financial accounts, and other records. Discovery of already breached information pooled across multiple breaches is not counted as an event. We focus on characterizing *extreme data breaches*, which credibly reduces issues of reporting and data completeness, as large breaches tend to be profoundly consequential and difficult to suppress (witness Uber, 2017[16]).[17] As seen above, moderate breaches as a whole appear to form the bulk of the total cost, but their frequencies cannot be studied on the basis of such incomplete data alone.

## 4.1 Overview

To emphasize the heavy tailed nature of breach sizes, with current data, see Figure 1: one out of ten breaches containing more than ten thousand pieces of private information (ids) will exceed one million ids. Statistics also differ by state, due to an indeterminate mix of varying factors at all levels of the framework, as well as in reporting behavior.[18] The cumulative total breach amounts to $1.4x10^{10}$ ids, of which the largest, being $3x10^{9}$, is 30 percent of the total. The 14 bil. total breach for the US alone already impresses the importance of strong protection of large exposures and vigorous mitigation in event of breach. Further precaution is warranted given that the full social cost of breaches are poorly known[19].

[14] “Expert Workshop on Improving the measurement of digital security incidents and risk management”. OECD. 12-13 May 2017. http://www.oecd.org/sti/ieconomy/improving-the-measurement-of-digital-security-incidents-and-risk-management.htm
[15] See http://www.privacyrights.org/data-breach.
[16] Uber Paid Hackers to Delete Stolen Data on 57 Million People, By Eric Newcomer, 21 November 2017 https://www.bloomberg.com/news/articles/2017-11-21/uber-concealed-cyberattack-that-exposed-57-million-people-s-data
[17] To test empirically, one could look at data breaches in each state separately and see if there was a change-point when a new cyber law was introduced. This could be an interesting way forward but goes beyond the current paper.
[18] Highest frequencies are in New York and California, with highest severities in Nebraska, Nevada, and D.C. See Appendix A1.

In the assessment, five breach event types are available (below), providing rough but still useful differentiation. More precise and mutually exclusive taxonomies exist but require application.[20]

- **HACK**: is any unauthorized exfiltration of ids by an outsider, typically including software media, and a range of attack strategies, excluding physical theft of devices.
- **HW**: is for all physical devices, i.e., hardware, either lost or stolen.
- **DISC**: is accidental disclosure via software media.
- **INSD**: is a HACK performed by an insider.
- **NA**: is not further specified, unknown, and/or does not fall into the above categories.

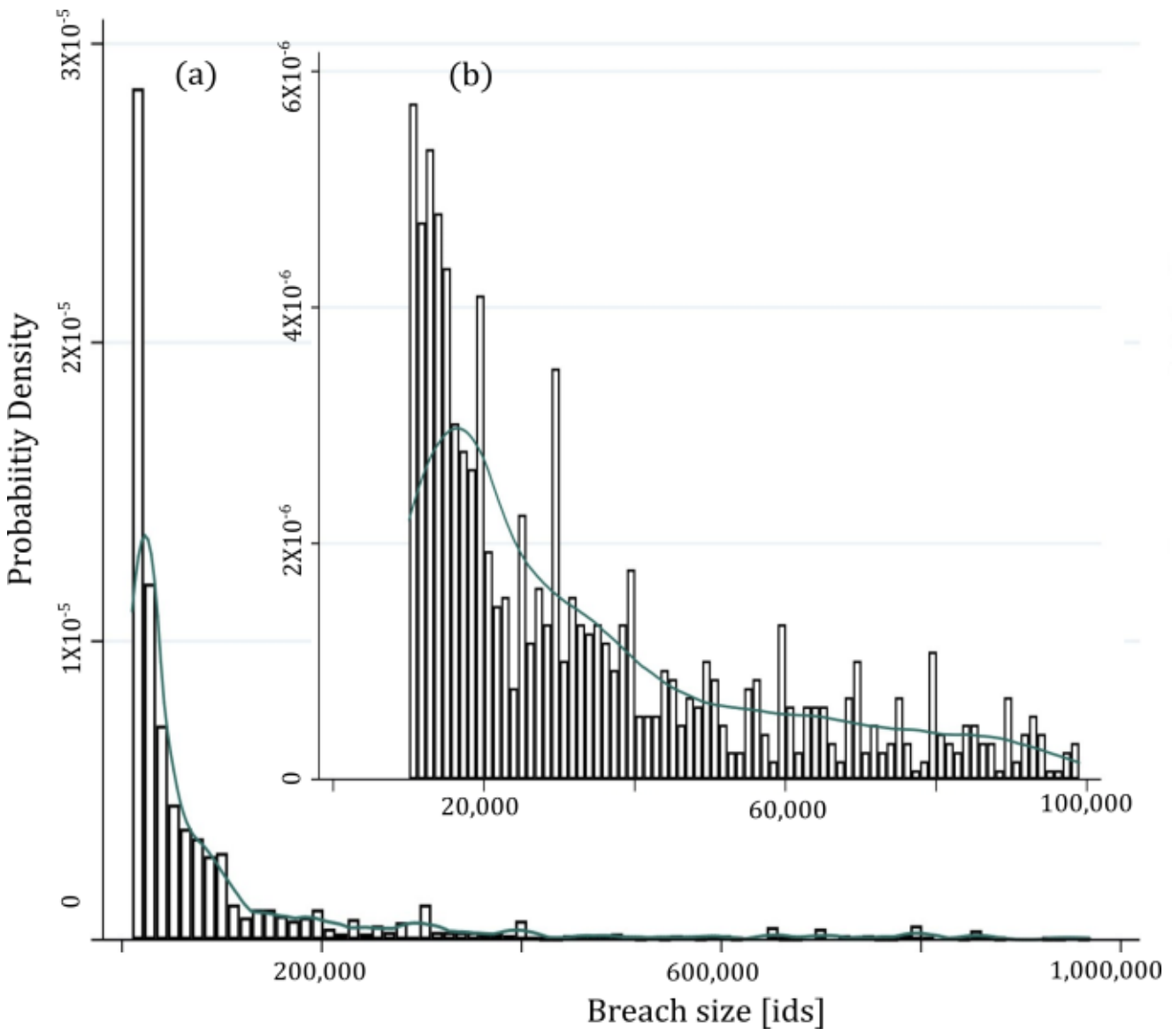

**Figure 1.** Histogram of cyber events up to (a) 1,000,000 and (b) 100,000 ids compromised.

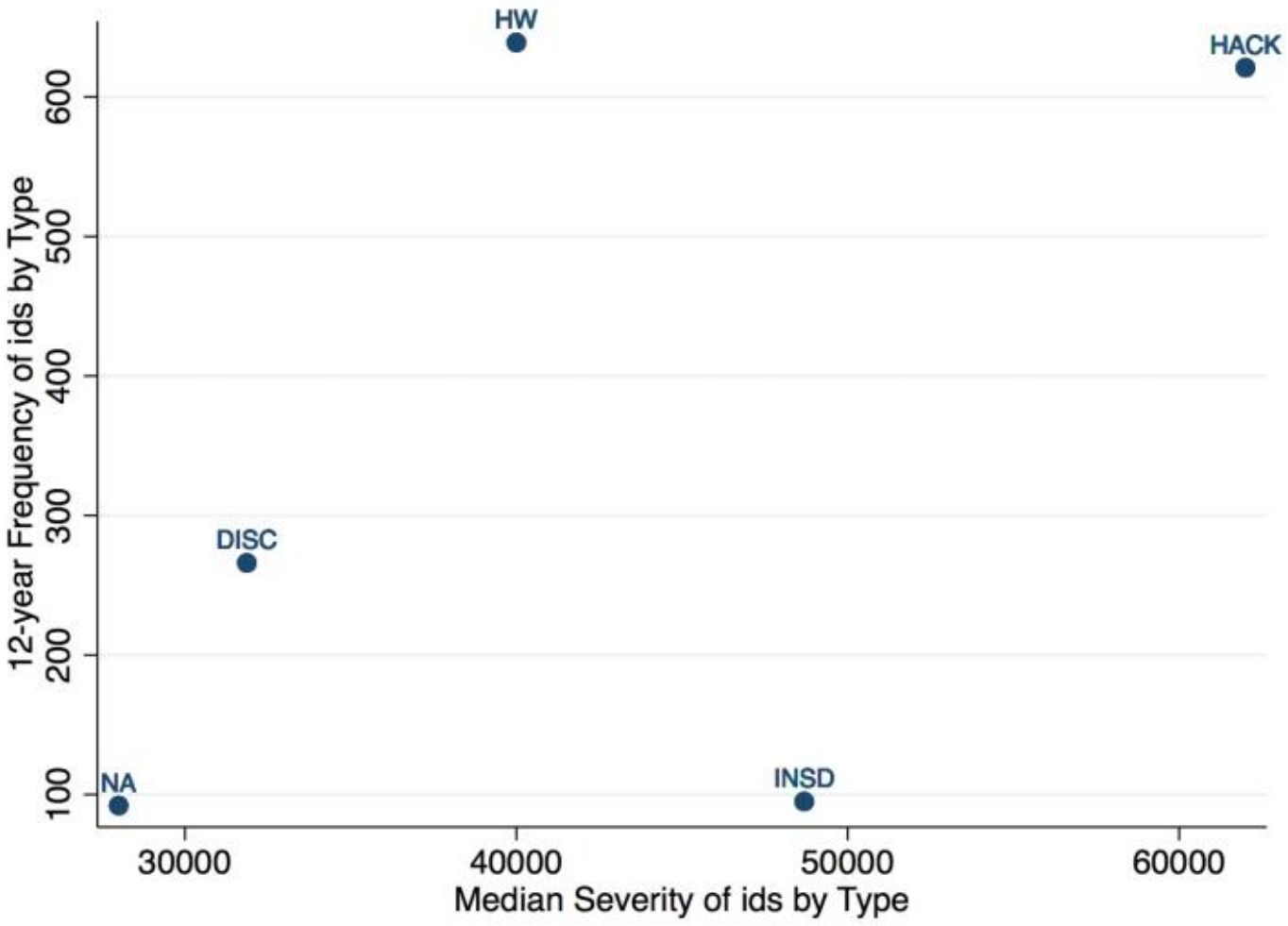

**Figure 2.** 12-year frequency and median severity by type of breach.

We let victim organizations be grouped into sectors: 1. financial, 2. non-financial business from which we distinguish 3. large web-based businesses/services, and social networks/communities, 4. medical, 5. educational, and 6. governmental entities. Summary statistics by sector and breach type are given in Tables 1 and 2, and Figure 2, again lacking normalization. Hack events tend to be the largest (especially at web-based companies) and most frequent (especially in the business sector). However, accidental disclosure is also significant, many of which warrant further study as precursors to more severe exploitation events.

## 4.2 Reporting delay

[19] Stolen identities have been used for fake comments online, distorting the appearance of important dialogues, and "hacking consensus". See, for example, the information on Hackernoon: *https://hackernoon.com/more-than-a-million-pro-repeal-net-neutrality-comments-were-likely-faked-e9f0e3ed36a6*

[20] Useful categories could include: External/internal actor, data media (hard or software), attack strategy/mode, intentional or accidental, actual effect or potential (i.e., precursor), data type, aggregating factors (e.g., distributed online), "cost" (e.g., total fraud, total liability, etc.).

Before analyzing the frequency and severity statistics, we get a sense for the effect of temporal incompleteness on the dataset. Indeed, for an event to be included, it needs to be discovered, reported, and curated into the database. Ponemon (2018) estimate that it takes about 196 days on average to identify a breach, whereas Mandiant (2015) find 205 days. An example of delay from discovery to reporting is the breach of 57 million ids from Uber in October 2016, which became known to the public in November 2017.[21] To approximate the full delay from event to its curation, data available to us are historical delays between event data and approximate submission date from Risk Based Security, when it was open ('datalossdb.com'). Considering years 2011 and 2012 (the most recent available), there are 698 non-zero size breach events, for US events in the business sector. This is visualized in Figure 3. Bursty arrival of submissions is natural – e.g., adding a new batch of events. At the same time, it can be an artefact of retroactively completing a database. Indeed for events submitted prior to 2010, the average event occurred about 550 days before being submitted. In the last half of 2012, with 214 submitted events, this had dropped to 190 days, with empirical quantiles: 25%: 18, 50%: 49 (median), 75%: 155, 90%: 570. We refer below to this empirical distribution as the delay distribution. Perhaps surprisingly, no apparent relationship was found between breach size and delay, as there was no significant slope in a linear regression of the log-transformed variables.

| **Quantile** | **50%** | **75%** | **90%** | **100%** | **Annual Freq.** |
|---|---|---|---|---|---|
| **Business** | 60,000 | 200,000 | 867,800 | 250.0 $x10^6$ | 26.1 |
| **Educational** | 34,000 | 80,750 | 185,100 | 7.5 $x10^6$ | 18.7 |
| **Financial** | 90,881 | 609,301 | 3,5 $x10^6$ | 145.5 $x10^6$ | 19.2 |
| **Government** | 72,000 | 250,000 | 1.3 $x10^6$ | 76.0 $x10^6$ | 12.4 |
| **Medical** | 29,082 | 79,750 | 327,359 | 78.8$x10^6$ | 53.2 |
| **Large Web** | 6.0$x10^6$ | 57$x10^6$ | 150$x10^6$ | 3.0$x10^9$ | 5.1 |

**Table 1.** Annual breach frequency for events in excess of 10'000 ids, and quantiles, by sector for firms in the US.

| **Type \ Sector** | **Business** | **Edu.** | **Financial** | **Gov.** | **Medical** | **Large Web** | **Sum (Bil.)** |
|---|---|---|---|---|---|---|---|
| **HACK** | 5.04$x10^8$ | 1.18$x10^7$ | 5.17$x10^8$ | 6.18$x10^7$ | 1.39$x10^8$ | <u>5.70$x10^9$</u> | <u>7.9</u> |
| **DISC** | 2.45$x10^8$ | 2.21$x10^6$ | 2.22$x10^6$ | 3.25$x10^7$ | 9.15$x10^6$ | <u>1.46$x10^9$</u> | <u>1.8</u> |
| **INSD** | 7.48$x10^6$ | 1.21$x10^6$ | 4.18$x10^7$ | 3.33$x10^7$ | 4.20$x10^6$ | 0 | 0.088 |
| **HW** | 2.33$x10^7$ | 2.80$x10^6$ | 3.62$x10^7$ | 1.31$x10^8$ | 5.89$x10^7$ | 2.80$x10^5$ | 0.25 |
| **Sum (Bil.)** | 0.78 | 0.028 | 0.61 | 0.26 | 0.21 | <u>7.3</u> | 10.4 |

**Table 2.** Total breached ids at U.S. firms, by sector and breach type, since 2005. Due to data limitations: a hack breach of 1 Billion ids in 2014 is not attributed to a sector, and overall 0.38 Billion are not attributed to an event type.

## 4.3 Frequency: are breaches becoming more frequent?

[21] Uber Breach, Kept Secret for a Year, Hit 57 Million Accounts. The New York Times. November 22, 2017. See https://www.nytimes.com/2017/11/21/technology/uber-hack.html

In Wheatley et al. (2016), it was claimed that the rate of breaches in excess of 50k ids was constant, and this continues to be the case. That is, GLM (generalized linear model) regressions give an insignificant slope over time. However, the most recent two years have only suffered about 80% of the historical mean frequency of events. This is likely to be a spurious decrease, consistent with the reporting delay distribution just discussed. For simplicity, we use smooth regression models that are largely robust to this incompleteness, although censoring methods could be applied.

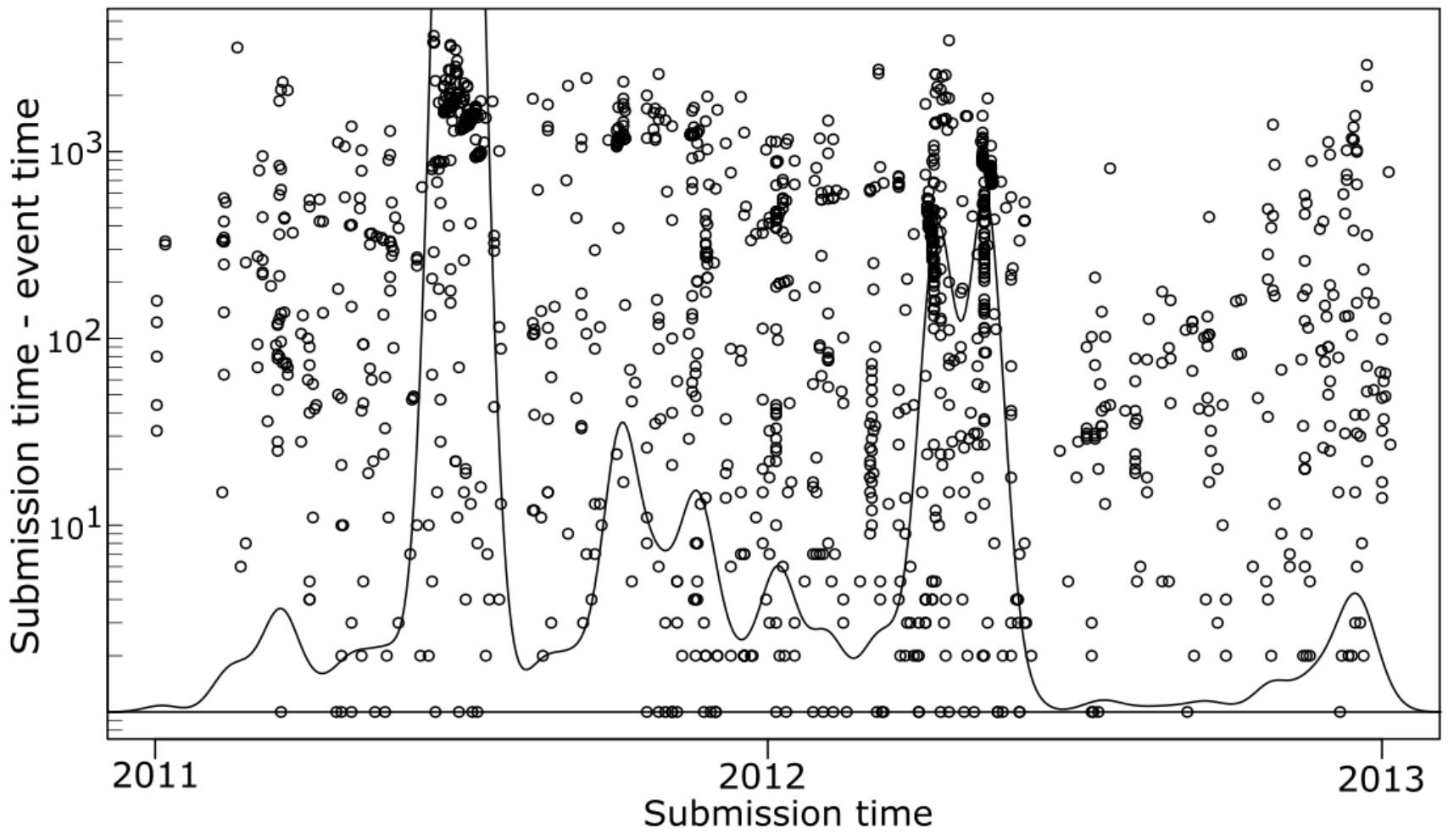

**Figure 3.** Delay between submission and event time for years 2011 and 2012, at the Open Security Foundation datalossdb. Plotted in logarithmic scale. The smooth line is a kernel density estimator of the intensity of submission activity, plotted in normal scale, without axis.

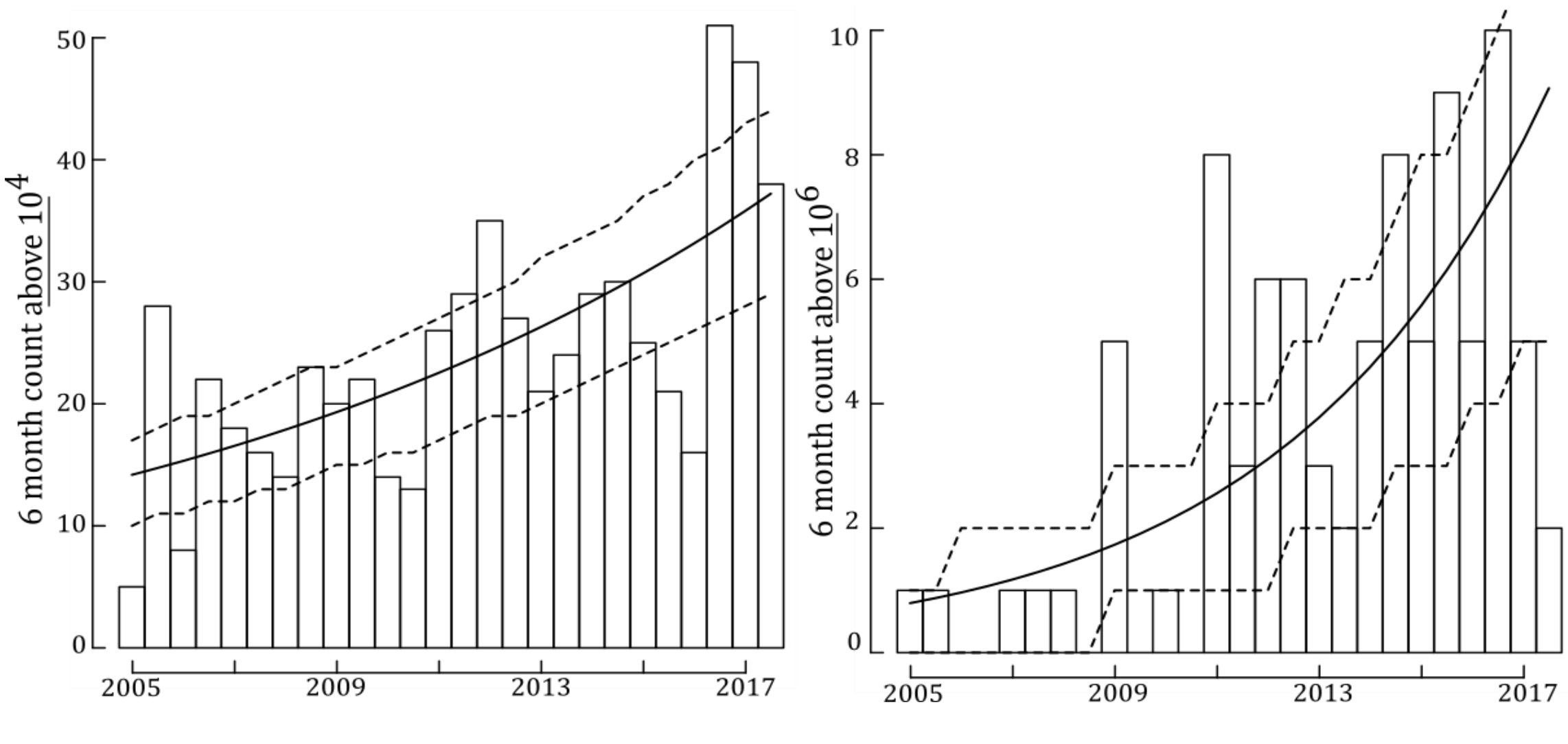

**Figure 4.** 6-month counts of hacking events from 01/2005 to 9/2017 (left with size above $10^4$ ids, right with size above $10^6$ ids). The dashed lines provide the quartiles of the estimated negative binomial distribution, roughly consistent with the fluctuations observed around the regression line.

On a more specific basis, looking at hack type events, we find a significant increase quantified by a log-linear negative binomial regression (see Figure 4 and Table 3). Annual growth ranges from 8 percent for breaches in excess of $10^4$ ids up to 19 percent per year for breaches larger than $10^6$ ids, having significantly faster growth for larger breaches.[22] The distribution of counts at any given time is well described by a negative binomial, with mean μ, and variance $\sigma^2 = \mu + \mu^2/\theta$. The fit is significantly over-dispersed relative to the Poisson: e.g., for the $10^4$ threshold, the variance-to-mean ratio grows from 2 to 3.5 over the almost 13 years of the dataset.

| **u** | **n>u** | **Intercept $\beta_0$** | **Growth rate $\beta_1$** | **θ** | P |
|---|---|---|---|---|---|
| $10^4$ | 623 | 2.65 (0.13) | 0.08(0.02) $10^{-5}$ | 14.8(7.0) 0.0003 | 0.28 |
| $10^5$ | 242 | 1.45 (0.19) | 0.11(0.02) $10^{-6}$ | 11.9(7.5) 0.05 | 0.38 |
| $10^6$ | 88 | -0.23 (0.36) | 0.19(0.04) $10^{-5}$ | 4.6(3.3) 0.15 | 0.21 |
| $10^7$ | 41 | -0.79 (0.48) | 0.17(0.06) .001 | 3.2(2.8) 0.47 | 0.35 |

**Table 3.** Log-linear negative binomial regressions (log link GLM fit by maximum likelihood), as a function of years 0 to 12.75 (01/2005 to 9/2017). Table quantities are: 1) lower threshold (u), 2) number of points above it, 3) estimated intercept with SE (standard error), 4) growth rate with SE and p-value for test with null being $\beta_1=0$, all of which are highly significant, 5) θ is the negative binomial dispersion parameter with SE and likelihood ratio test p-value against Poisson null, indicating significant over-dispersion, and 6) "p" is the chi-square test p-value on deviance residuals (model diagnostic), indicating that the overall fits cannot be rejected.

### 4.4 Severity: are big breaches getting bigger?

As demonstrated above, there is a strong trend towards increasing severity of particularly high-impact cyber events in our data, which is not present for all cyber events. Another approach is to examine the full event size distribution over time. As can be seen in Figure 5, one can distinguish a clear and significant increasing trend for hack events – consistent with the regressions in Table 3, whereas other event types are roughly stationary and less extremal.

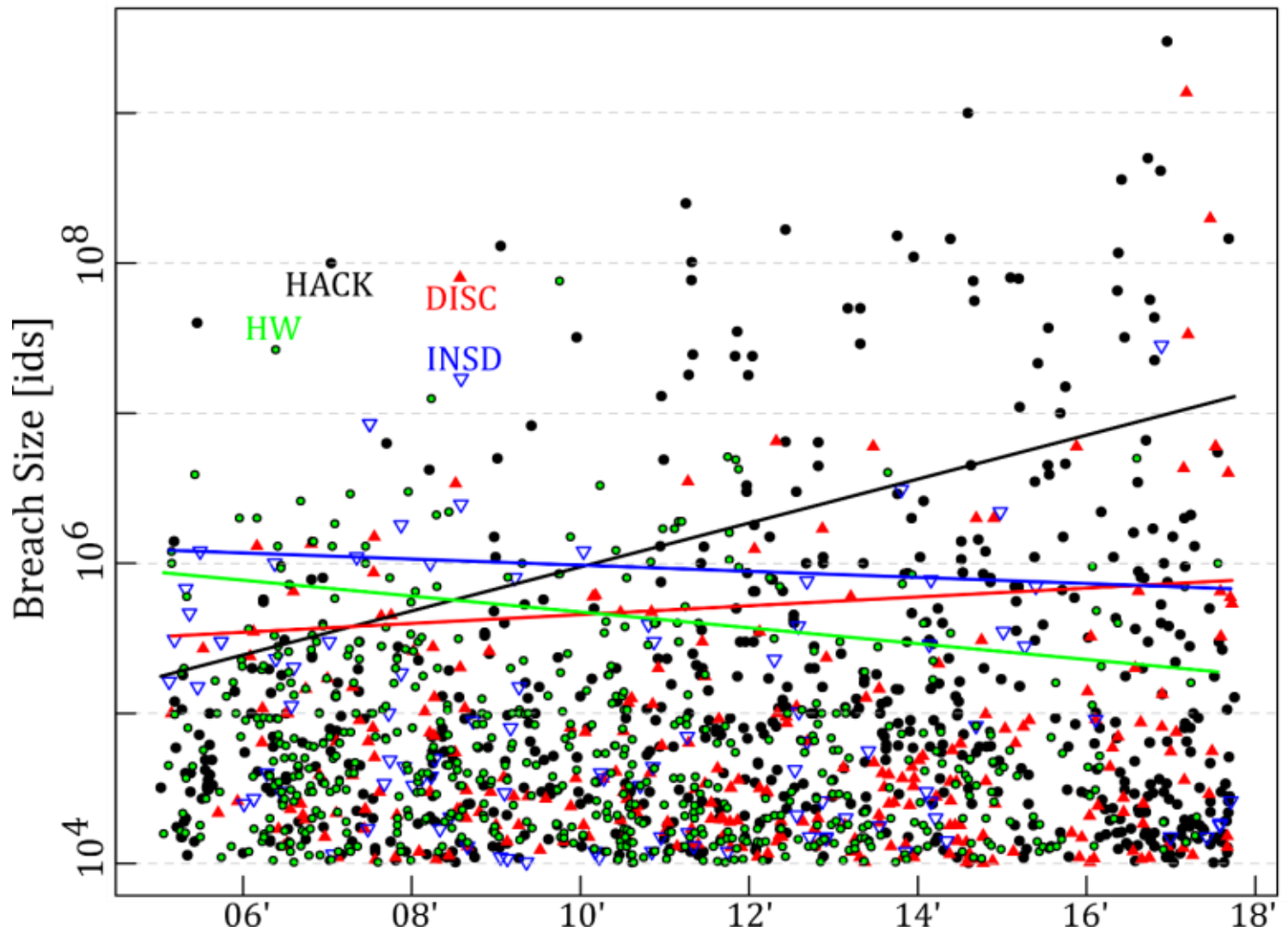

[22] One sided Z test of the growth rates of the $u=10^6$ and $u=10^4$ fits gives p=0.007, indicating significantly faster growth of more extreme breaches.

**Figure 5.** Scatterplot of breach sizes over the years, by type (HACK black, DISC red, INSD blue, HW green). The trend is clear for hacks but not for others, as quantified by the log-linear regression of the 0.9 quantile (Koenker, 2001). For the HACK data, the slope is significant with p-value < 0.01.

Without assuming a specific dynamic model for the growth of hacking events, the Pareto distribution

$$Pr\{X > x\} = (x/u)^{-\alpha}, x > u > 0, \alpha > 0 \quad (1)$$

is fitted on a moving window of 50 events to capture the non-stationary behavior. As shown in Figure 6, a major decrease of the tail parameter α took place, from the range 0.6-0.7 before 2007 to a roughly stable value in the range 0.3-0.4 post 2015, as the tail has become heavier and significantly bent-down relative to a pure Pareto, for the largest values. To consider alternative tail models, a lower-truncated lognormal[23] and an upper-truncated Pareto are also considered, where the truncated CDF (cumulative distribution function),

$$T(x) = Pr\{X \le x | u < X \le m\} = \frac{F(x)-F(u)}{F(m)-F(u)}, 0 < u < m, T(u) = 0, T(m) = 1\,, \quad (2)$$

is derived from its un-truncated CDF, F(x). The tail fits shown and summarized in Table 4 indicate that both a lognormal and an upper-truncated Pareto[24] tail fit well, and are statistically indistinguishable.[25] Naturally, a maximal breach size exists, on the order tens of billions, where breaches in excess of 1 billion ids have already occurred. Given trends already discussed, this maximum is likely to grow. In this case, only the upper truncated Pareto imposes a finite maximum.

| u | N>u | α | logL | $\alpha_1$ | logL | μ | $\sigma^2$ | logL |
|---|---|---|---|---|---|---|---|---|
| 25k | 167 | 0.35(.03) | -341 | 0.3(.03) | -338 | -12.8(13) | 7.1(3) | -340 |
| 100k | 101 | 0.35(.03) | -208 | 0.3(.03) | -204 | -2.9(4.0) | 4.7(1.3) | -205 |
| 1mil. | 47 | 0.40(.05) | -90 | 0.3(.05) | -87 | -4.1(7.4) | 4.5(2.1) | -89 |

**Table 4.** Tail fits of hack breach sizes since 2014 with Pareto (left), upper-truncated Pareto (middle, parameter $\alpha_1$), and Lognormal distributions (right). The lower threshold (u), maximum likelihood parameter estimate, standard error, sample size, and log-likelihood value are given for all fits.

## 4.5 Risk: An aggregate compound process

Focusing on the dominant and worsening breach risk, caused by hack events, we propose a model for the U.S. in the near future (with analysis done in Q2 of 2018). The incidence is well modeled by a highly dispersed negative binomial distribution with exponentially growing mean (Table 3). However, a linear mean model is not significantly worse,[26] which yields lower future predicted values. For severity, a Pareto distribution with lower bound $u=10^4$, maximum value $m=10^{10}$ and estimated $\alpha = 0.35$ is proposed – being between the best Pareto and lognormal fits (see Figure 6). Accounting for model and parameter uncertainties, the aggregate distribution is summarized, for the last 6 months of 2018, in Table 5. This predicts a median of around 0.5 Billion hacked ids in the last 6 months of 2018 of the US, but allows for more than 7 bil. ids to be breached with about a 5% chance – which would about double the historical breach due to hacks (see Table 2). Despite the imposed finite maximum, the tail is *exceptionally heavy*, resulting in the potential for massive fluctuations in aggregate loss. As it turned out, 'fortunately', the observed breach in Q3 and Q4 of 2018 thus far is about 4.5 bil. – close to the median predicted. However this is not yet a final figure, as the empirical delay distribution discussed in section 4.2 indicates that, as of 1-5-2019, the Q3 and Q4 period of 2018 is expected to eventually contain 20% more events – and those being drawn from heavy tailed breach size distribution.

[23] The distribution such that the natural logarithm is the lower-truncated Normal distribution with parameters (μ, $\sigma^2$). See Malevergne et al. (2011) for examples, and the uniformly most powerful unbiased test against the Pareto tail.
[24] Pareto distribution with upper truncation set to the size of the largest HACK event: $3\times10^9$.
[25] According to a chi-square likelihood ratio test, only the un-truncated Pareto is significantly worse ($p < 0.01$).
[26] With intercept 13.1 (2.3), slope 1.7 (0.4), and NB dispersion parameter 14.3 (6.65), e.g., predicting a level of 13.1 events per 6 months in 2005, and approx. 35.2 by Q3 2017, where the exponential model predicts 38.4.

These figures for the Q3 and Q4 2018 are briefly compared with the aggregate distribution for the Q3 and Q4 of 2012, summarized in Table 6. In particular, $\alpha = 0.4$ is selected based on Figure 6, a maximum breach event of $m = 10^9$ is assumed, and frequency is again taken according to the GLM model. On this basis, the mean and quantiles are about ten times less than their counterparts in Table 5, quantifying an *order of magnitude worsening of risk over the past five years*.

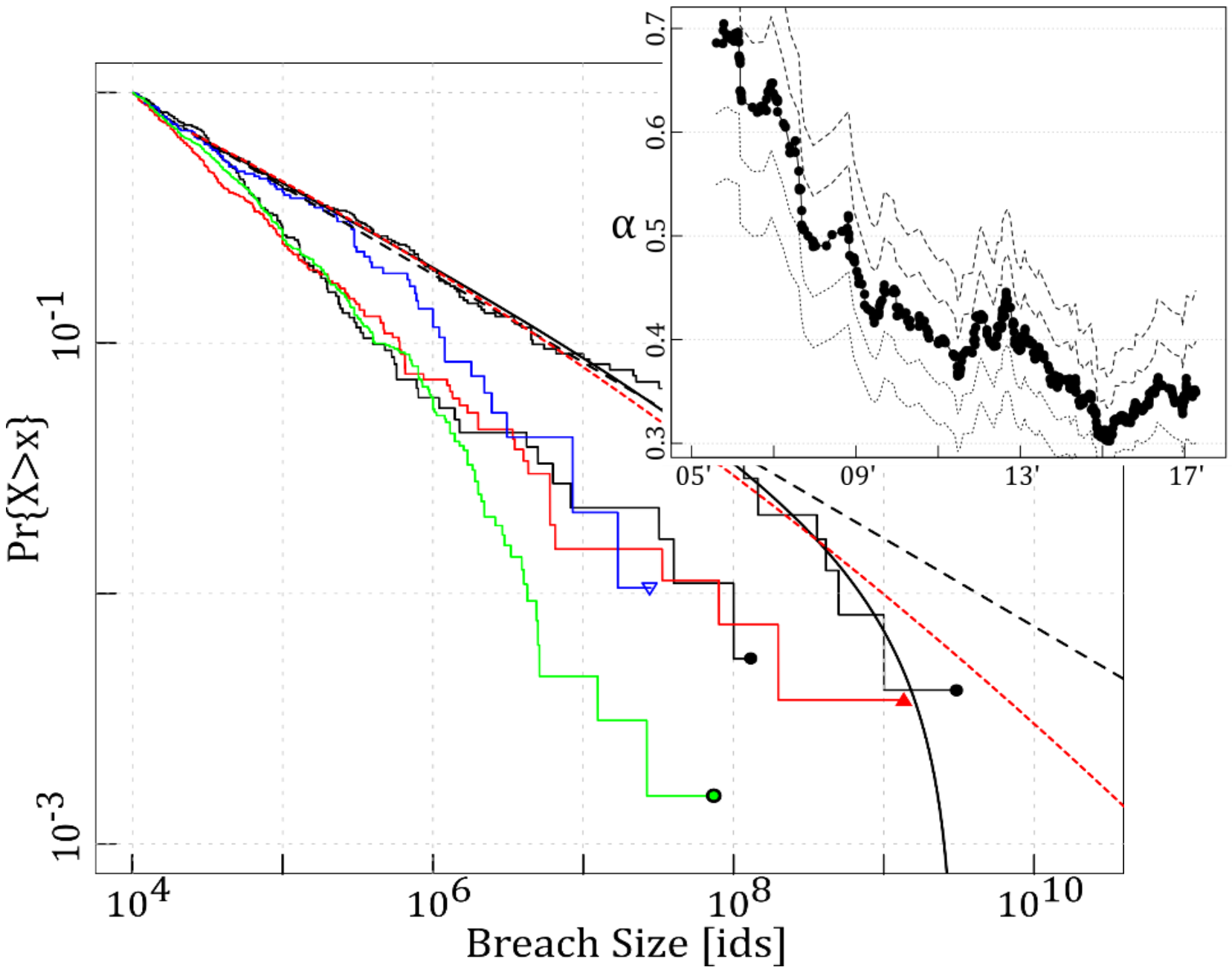

**Figure 6**: Main: Empirical survival distributions for the breach types for all time by color: DISC in red, INSD in blue, and HW in green. HACK (in black) is split into pre-2010 data and post-2014 data, with the latter being the heavier tailed. Maximum likelihood fits to the post-2014 hack data are plotted: Pareto with upper truncation (black solid), with no truncation (black dashed), and Lognormal (red dashed). Inset: Estimated parameter α of the Pareto tail (u=25'000), from 2005 to Q4 2017 on a moving window of 50 points, for the HACK events, including one and two standard deviations of the maximum likelihood estimate.

| Quantile | Mean | SD | 0.5 | 0.9 | 0.95 | 0.99 |
|---|---|---|---|---|---|---|
| 0.25 | 1.5 | 2.5 | 0.41 | 4.9 | 7.3 | 11 |
| 0.5 | 1.6 | 2.6 | 0.5 | 5.4 | 7.8 | 11.4 |
| 0.75 | 1.9 | 2.8 | 0.61 | 5.8 | 8.2 | 12.1 |

**Table 5:** Summary of aggregate distribution for total number of billions of ids hacked in the last six months of 2018 (out of sample at time of writing with data sample studied ending at 09/2017). A lower threshold of ten thousand was taken. The mean, standard deviation, and tail quantiles (VaR) are given. The rows are the median and quartiles of these quantities accounting for model uncertainty (equal weighting for the exponential and linear mean model for frequency) and parameter uncertainty (by parametric bootstrap).

| Quantile | Mean | SD | 0.5 | 0.9 | 0.95 | 0.99 |
|---|---|---|---|---|---|---|
| 0.25 | 0.18 | 0.26 | 0.06 | 0.54 | 0.77 | 1.11 |
| 0.5 | 0.19 | 0.27 | 0.07 | 0.57 | 0.81 | 1.18 |
| 0.75 | 0.20 | 0.28 | 0.08 | 0.61 | 0.85 | 1.25 |

**Table 6:** The same as table 5, also given in billions of ids, but for the in-sample period of the first 6 months of 2013.

## 5. Conclusions

Considering cyber risk within the cat-modelling framework, as a man-made catastrophe, provides guidelines for the assessment and quantification of cyber risk, useful for insurance. The top-down statistical operational risk perspective was contrasted and complemented with the more bottom-up modelling helpful for operational safety/security and refinement. All of the above provide some ideas for which advances in data and modelling are necessary, and hints for how to proceed. The future role of data collection and evaluation was emphasized. The important and relatively tractable 'damage' level of the framework was addressed. In particular, with open data for US firms, it was found that 'hack' type data breach events are becoming increasingly heavy tailed – well described by an extremely heavy tailed truncated Pareto, or a lognormal. Characterization of the other modules will enable additional and more complete and useful insights. Hopefully, this provides a step towards reducing ambiguity and enabling maturation of cyber insurance cover – potentially triggering higher protection within the whole network (i.e., by tipping point) as a positive externality.

## References

Ayoub, A., Kröger, W., Nusbaumer, O., and Sornette, D. (2019), Simplified/harmonized PSA: a generic modeling framework applied to precursor analysis, PSA 2019, 16th International Topical Meeting on Probabilistic Safety Assessment and Analysis, Charleston, South Carolina

Betterley, R. (2013), Cyber/Privacy Insurance Market Survey 2013: Carriers Deepen Their Risk Management Services Benefits - Insureds Grow Increasingly Concerned with Coverage Limitations, online edition, 2013, available online at: *http://betterley.com/samples/cpims13_nt.pdf.*

Boehme, R., Laube, S., and Riek, M. (2017). A Fundamental Approach to Cyber Risk Analysis, Variance Journal, www.variancejournal.org, online edition, 2017. Available online at: http://www.variancejournal.org/articlespress/articles/Fundamental-Boehme.pdf

Biener, C., Eling, M. and Wirfs, J. H. (2015). Insurability of Cyber Risk: an Empirical Analysis. Geneva Papers on Risk and Insurance-Issues and Practice, 40(1):131-158.

Bouchaud, J.-P., D. Sornette, C. Walter and J.-P. Aguilar (1998). Taming large events : Optimal portfolio theory for strongly fluctuating assets, International Journal of Theoretical and Applied Finance 1: 25-41.

Cebula, J.J., Popeck, M.E., and Young, L.R. (2010). A taxonomy of operational cyber security risks, Technical Note, CMU/SE-2010-TN-028. Software Engineering Institute, Carnegie Mellon University.

Chernov, D. and Sornette, D. (2016). Man-made catastrophes and risk information concealment (25 case studies of major disasters and human fallibility), Springer, 1st ed. 2016 edition.

Chernov, D. and Sornette, D. (2019). Critical risks in different economic sectors (more than 500 case studies), Springer (in press).

Cisco (2017). Midyear Cybersecurity Report. online edition, 2017. Available online at: https://engage2demand.cisco.com/LP=5897.

CRO Forum (2018). Emerging Risks Initiative : Major Trends and Emerging Risk Radar April 2018 Update, CRO Forum. https://www.thecroforum.org/wp-content/uploads/2018/05/CRO-ERI_Emerging-Risk-RadarTrends_Apr2018_FINAL.pdf

EDPB (2019), First overview on the implementation of the GDPR and the roles and means of the national supervisory authorities, European Union. https://edpb.europa.eu/sites/edpb/files/files/file1/19_2019_edpb_written_report_to_libe_en.pdf

Edwards, B., Hofmeyr, S. and Forrest, S. (2016). Hype and heavy tails: A closer look at data breaches, Journal of Cybersecurity, 2(1):3-14.

Eling, M., and Schnell, W. (2016a). What do we know about cyber risk and cyber risk insurance? Journal of Risk Finance, 17(5):474-491.

Eling, M., and Schnell, W. (2016b). Ten Key Questions on Cyber Risk and Cyber Risk Insurance, Geneva Association Newsletter, November 2016 Report. Available online at: https://www.genevaassociation.org/sites/default/files/research-topics-document-type/pdf_public//cyber-risk-10_key_questions.pdf.

Eling, M., and Schnell, W. (2018). Extreme Cyber Risks and the Nondiversification Trap, Working Paper, August 2018.

Eling, M. and Wirfs, J. H. (2015). Modelling and management of cyber risk. Available online at: https://www.actuaries.org/oslo2015/papers/IAALS-Wirfs&Eling.pdf.

Eling, M., and Wirfs, J.H. "What are the actual costs of cyber risk events?." European Journal of Operational Research 272.3 (2019): 1109-1119.

ENISA (2016). The Cost of Incidents Affecting CIIs. ENISA 2016. https://www.enisa.europa.eu/publications/the-cost-of-incidents-affecting-ciis

Esentire (2019). Nearly half of firms suffer data breach at hands of vendors, esentire.com https://www.esentire.com/blog/nearly-half-of-firms-suffer-data-breach-at-hands-of-vendors/

Europol (2016). Europol's 2016 Internet Organised Crime Threat Assessment (IOCTA). https://www.europol.europa.eu/internet-organised-crime-threat-assessment-2018 .

Gordon, L. A., and Sohail, T. (2003). A framework for using insurance for cyber-risk management. Communications of the ACM, 46(3):81-85.

Grossi, P., and Kunreuther, H. (2005). "Catastrophe modeling: a new approach to managing risk." Huebner international series on risk, insurance and economic security.

Hofmann A. and Ramaj (2011). Interdependent Risk Networks, *International Journal of Management and Decision Making,* Vol. 11, No. 5/6, 312-323, 2011.

Ibragimov, R., and Walden, J. (2009). Nondiversification Traps in Catastrophe Insurance Markets, Review of Financial Studies, 22(3): 959–993.

Ibragimov, R. and Walden, J. (2007). The Limits of Diversification when Losses may be Large, Journal of Banking & Finance, 31(8): 2551–2569.

Jacobs (2014). Analyzing Ponemon Cost of Data Breach, datadrivensecurity.com. https://datadrivensecurity.info/blog/posts/2014/Dec/ponemon/

Kaplan, S. and Garrick, J. (1981). On the quantitative definition of risk, *Risk Analysis*, 1(1): 11-27.

Kessler (2018). Cyber Risk Survey Report 2018 – Cyber Risk from a Swiss Perspective. https://www.kessler.ch/fileadmin/user_upload/KS_Cyber_Report_2018_EN.pdf

Koenker, R. and Hallock, K.F. (2001). Quantile regression. Journal of Economic Perspectives, 15(4), pp.143-156.

Kovalenko, T. and Sornette, D. (2016). Risk and Resilience Management in Social-Economic Systems, published in the IRGC Resource Guide on Resilience (2016) (https://www.irgc.org/irgc-resource-guide-on-resilience), (http://ssrn.com/abstract=2775264)

KPMG (2016). Small Business Reputation & The Cyber Risk, KPMG. https://home.kpmg/content/dam/kpmg/pdf/2016/02/small-business-reputation-new.pdf

Kröger W. (2019) Achieving Resilience of Large-Scale Engineered Infrastructure Systems. In: Noroozinejad Farsangi E., Takewaki I., Yang T., Astaneh-Asl A., Gardoni P. (eds) Resilient Structures and Infrastructure. Springer, Singapore

Kumar, V., Telang, R. and Mukhopadhyay, T. (2007) Optimally securing interconnected information systems and assets, *Proceedings of the Sixth Workshop on the Economics of Information Security*, 7–8 June,

Carnegie Mellon University. Available online at *http://citeseerx.ist.psu.edu/viewdoc/summary?doi=10.1.1.209.425.*

Kunreuther, H. and Heal, G. (2003). Interdependent security, *Journal of Risk and Uncertainty*, Vol. 26, pp.231–249.

Kunreuther, H., Bier, V.M. and Phimister, J.R. eds., (2004). Accident precursor analysis and management: reducing technological risk through diligence. National Academies Press.

Leveson, N. (2011). Engineering a safer world: Systems thinking applied to safety. MIT press.

Lloyd's (2017). Counting the cost: Cyber exposure decoded, Emerging Risks Report 2017, Lloyds/Cyence.

Maillart, T. and Sornette, D. (2010). Heavy-tailed distribution of cyber-risks, *European Physical Journal B*, 75(3):357-364.

Maillart, T., Sornette, D., Frei, S., Duebendorfer, T. and Saichev, A. (2011) Quantification of deviations from rationality from heavy-tails in human dynamics, Physics Review E 83, 056101.

Malevergne, Y., Pisarenko, V., and Sornette, D. (2011). Testing the Pareto against the lognormal distributions with the uniformly most powerful unbiased test applied to the distribution of cities. Physical Review E, 83(3), 036111.

Marotta, A., et al. (2017) Cyber-insurance survey. Computer Science Review 24: 35-61.

Mukhopadhyay, A., Chatterjee, S., Saha, D., Mahanti, A. and Sadhukhan, S. K. (2013). Cyber-risk decision models: To insure IT or not? *Decision Support Systems*, 56:11-26.

Oeguet, H., Raghunathan, S. and Menon, N. (2011). Cyber security risk management: Public policy implications of correlated risk, imperfect ability to prove loss, and observability of self-protection, *Risk Analysis*, 31(3)

Opadhyay, T. B., V. S. M. and Rao, R. C. (2009). Why IT managers don't go for cyber-insurance products. *Communications of the ACM*, 52(11):68-73.

ORX (2018). Annual Insurance Risk Report, ORX. https://managingrisktogether.orx.org/sites/default/files/downloads/2018/07/annual_insurance_loss_report_2018.pdf

Ponemon (2017). 2014 Cost of Data Breach Study, IBM/Ponemon.

Ponemon (2018). 2017 Cost of Data Breach Study, IBM/Ponemon.

Proofpoint (2019). The latest in phishing: first of 2019. Proofpoint. https://www.proofpoint.com/us/security-awareness/post/latest-phishing-first-2019

RMS (2017). Cyber Risk Landscape, RMS, 2017. Available at: http://www.rms.com/models/cyber.

Romanosky, S., Ablon, L., Kuehn, A. and Jones, T. (2017). Content Analysis of Cyber Insurance Policies, Working Paper, RAND Justice, Infrastructure, and Environment, 2017.

Romanosky, S. (2016). Examining the costs and causes of cyber incidents. Journal of Cybersecurity 2.2: 121-135.

RSA (2018). Current State of Cybercrime, RSA. https://www.rsa.com/content/dam/premium/en/white-paper/rsa-2018-current-state-of-cybercrime.pdf.

Rothschild, M., Stiglitz, J. E. (1976). Equilibrium in Competitive Insurance Markets: An Essay on the Economics of Imperfect Information, *Quarterly Journal of Economics*, 90(4), 629-649.

Saichev, A., and D. Sornette (2010). Effects of diversity and procrastination in priority queuing theory: The different power law regimes. Physical Review E 81, 016108

Schelling, T. (1978) Micromotives and Macrobehavior, W.W. Norton and Firm, New York.

Shevchenko, N. et al. 2018. Threat Modeling: A summary of available methods, Software Engineering Institute, Carnegie Mellon University. https://resources.sei.cmu.edu/asset_files/WhitePaper/2018_019_001_524597.pdf

Shetty, S., McShane, M., Zhang, L., Kesan, J.P., Kamhoua, C.A., Kwiat, K., Njilla, L.L. (2018). Reducing informational disadvantages to improve cyber risk management, *Geneva Papers on Risk and Insurance – Issues and Practice*, forthcoming.

Sornette, D., Wolfgang K., and Wheatley, S. (2018). New Ways and Needs for Exploiting Nuclear Energy. Springer.

SRA (2015), SRA glossary. Society for Risk Analysis. (2015) https://www.sra.org/sites/default/files/pdf/SRA-glossary-approved22june2015-x.pdf

Stalder, I. (2017). Assessing man-made catastrophes, Zurich Insurance. https://www.zurich.com/en/knowledge/articles/2017/03/assessing-man-made-catastrophes

SwissRe (2017). Cyber: getting to grips with a complex risk. https://www.swissre.com/dam/jcr:995517ee-27cd-4aae-b4b1-44fb862af25e/sigma1_2017_en.pdf

Trump, B. D., Florin, M.-V., and Linkov, I. (Eds.) (2018). IRGC resource guide on resilience (vol. 2): Domains of resilience for complex interconnected systems. Lausanne: EPFL International Risk Governance Center (IRGC). Available from: irgc.epfl.ch

WEF (2018). The Global Risks Report 2018 13th Edition https://www.weforum.org/centre-for-cybersecurity

Wheatley, S., Maillart, T., Sornette, D. (2016). The extreme risk of personal data breaches and the erosion of privacy, *The European Physical Journal B*, 89(7):1-12.

## A.1 Cyber events in the United States by State

Figure A1 shows the 12-year frequency and median severity of information items compromised by U.S. state. As can be seen, the highest frequencies are in the States of New York and California, followed by Texas and Ohio. The highest severity, interestingly, has Nebraska, followed by Nevada and District of Columbia.

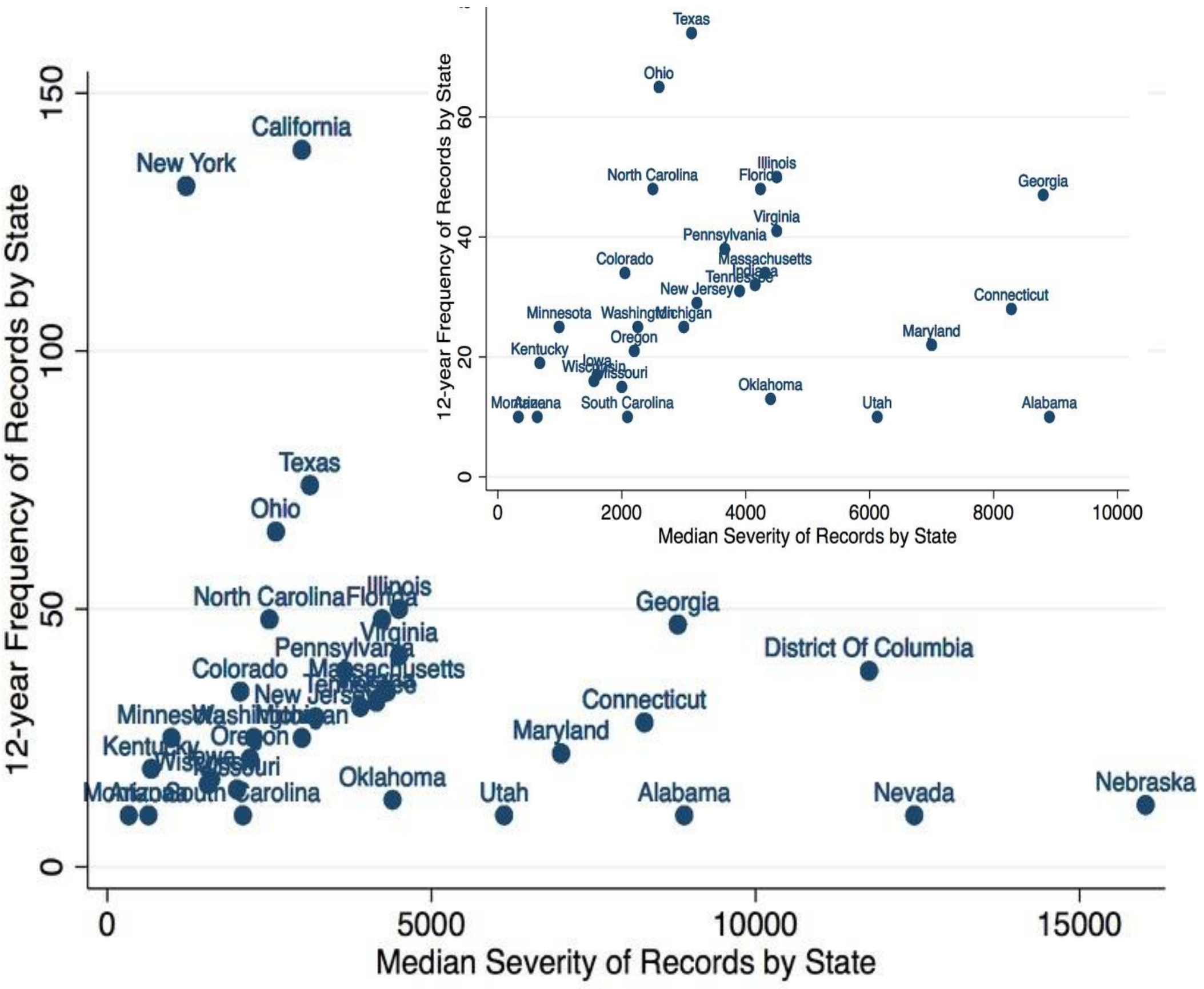

**Figure A1:** 12-year frequency and median severity of ids (information items compromised) by U.S. state. *Notes*: We use only breach events from the PRC [rather than OSF DLDB] database, where state information is readily accessible. States with ≤ 10 cyber incidents over the 12 years were omitted [i.e. Alaska, Arkansas, Delaware, Hawaii, Idaho, Kansas, Louisiana, Maine, Mississippi, New Hampshire, New Mexico, North Dakota, Rhode Island, South Dakota, Vermont, West Virginia, Wyoming]. The inset panel magnifies the cluster near the origin.

## A.2 Chronology of cyber events: Is there a general trend?

There is a strong trend for increasing frequency and severity of hack data breach events, having size > 10,000k. This very strong trend is not distinguishable for overall breach events in excess of 10k (see Figures A2 and A3).

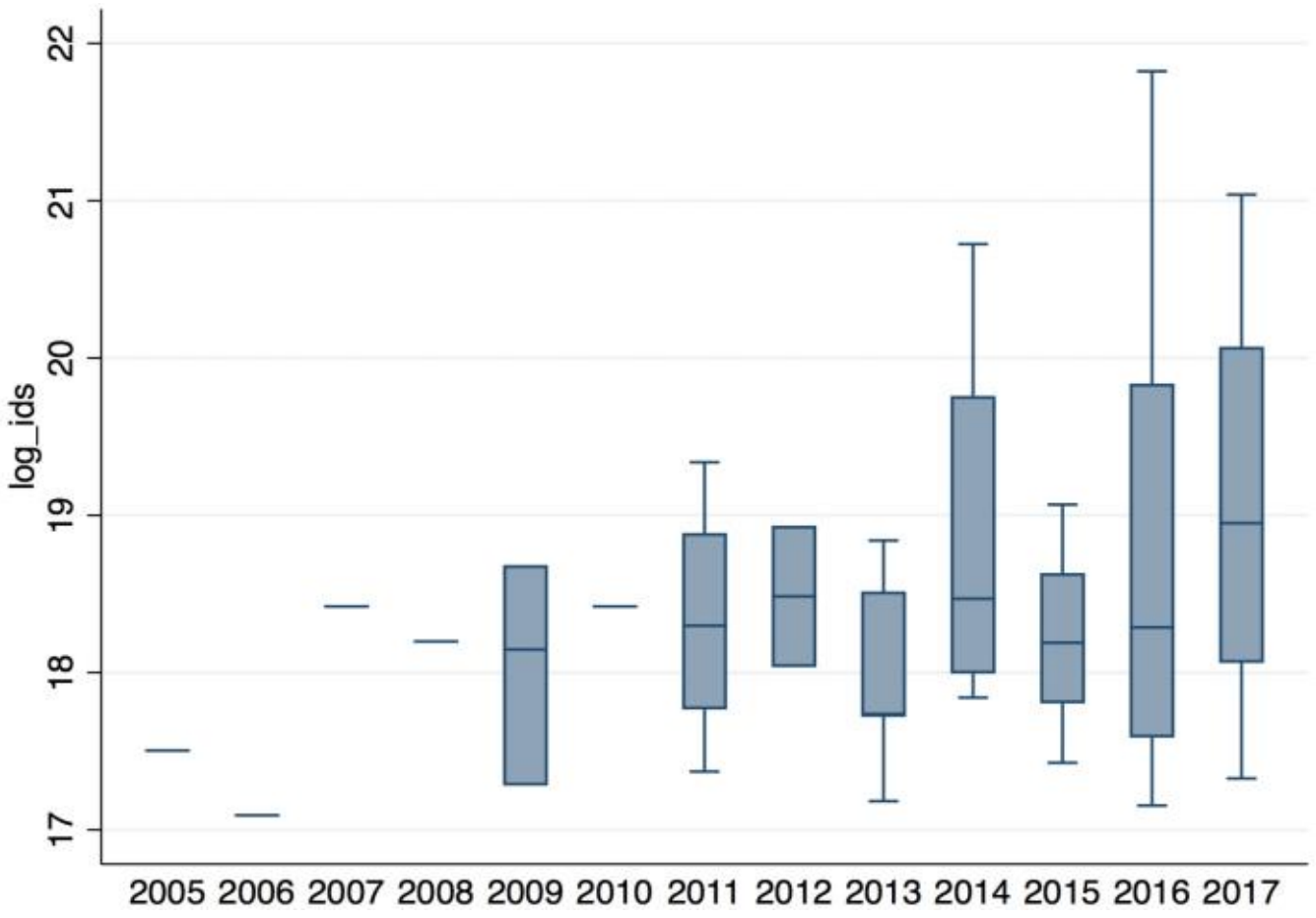

**Figure A2:** Chronology of high-impact cyber events involving ids > 25,000k compromised during 01/2005 through 9/2017. We can see a <u>strong</u> upward trend in severity here. *Note*: Natural logarithmic scale for ids along the ordinate axis.

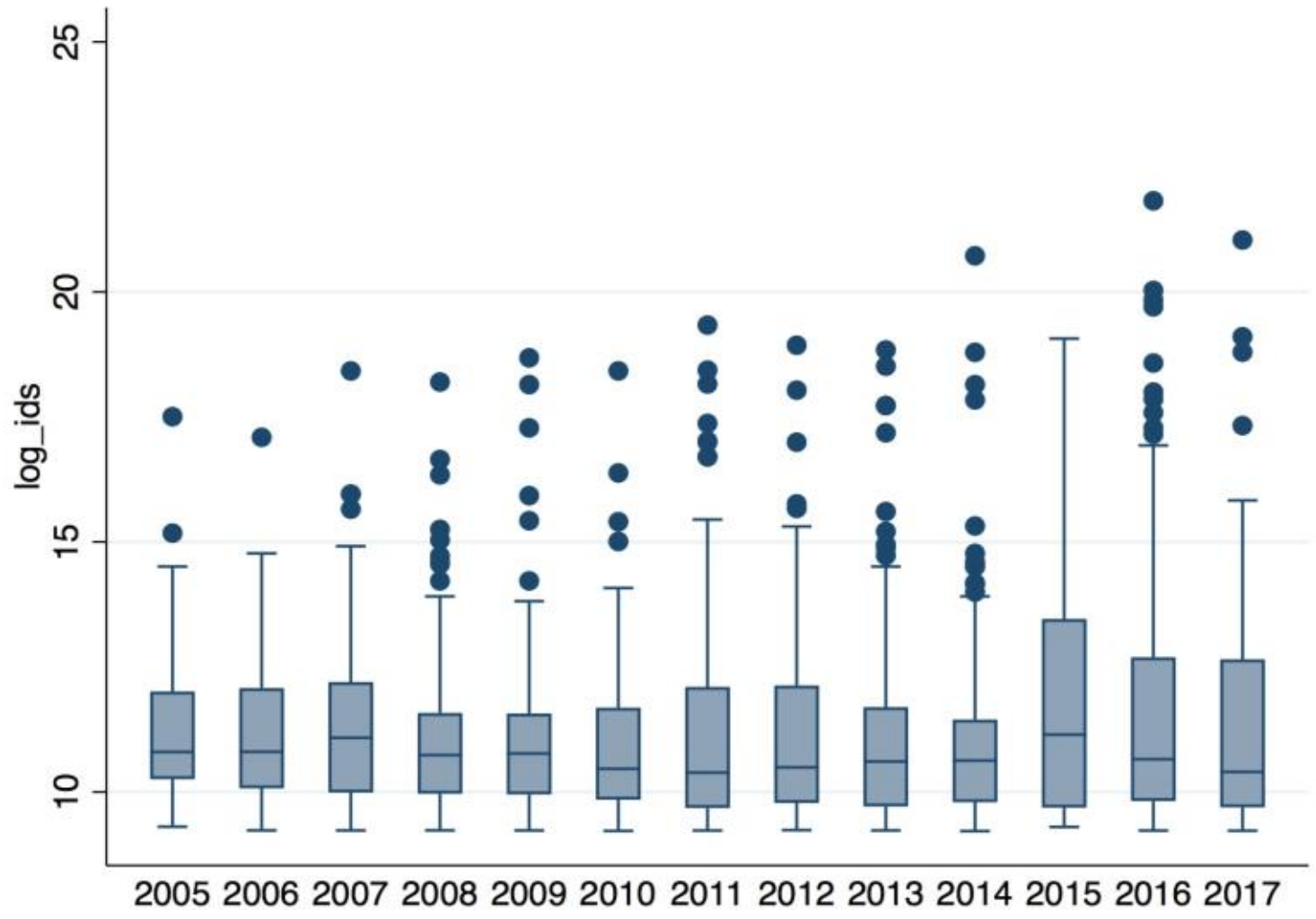

**Figure A3:** Chronology of <u>all</u> breach events from 01/2005 through 9/2017. As can be seen, the overall picture for the severity shows a <u>weak</u> upward trend. A more specific view (as shown in figure A2) is necessary to identify trends. *Note*: Natural logarithmic scale for ids.